\documentclass{article}
\usepackage{graphicx}
\usepackage{amsmath}
\usepackage[hidelinks]{hyperref}
\usepackage{tabularx,booktabs}
\usepackage{enumitem}
\usepackage{multirow}
\usepackage{array}
\usepackage{enumitem}
\usepackage{nameref}
\usepackage{adjustbox}
\usepackage{xcolor}
\usepackage{tikz}
\usetikzlibrary{positioning}

\usepackage{geometry}

\newlist{researchq}{description}{1}
\setlist[researchq,1]{labelwidth=\widthof{\bfseries RQ\ref{rqi}}, leftmargin=!}

\newcolumntype{P}[1]{>{\raggedright\arraybackslash}p{#1}}
\newcolumntype{M}[1]{>{\raggedright\arraybackslash\ttfamily}m{#1}}
\newcommand{\todo}[1]{}
\renewcommand{\todo}[1]{\pdfliteral{1 0 0 rg}#1\pdfliteral{0 0 0 rg}}

\makeatletter
\let\orgdescriptionlabel\descriptionlabel
\renewcommand*{\descriptionlabel}[1]{%
	\let\orglabel\label
	\let\label\@gobble
	\phantomsection
	\protected@edef\@currentlabel{#1\unskip}%
	\let\label\orglabel
	\orgdescriptionlabel{#1}%
}
\makeatother

\begin{document}
\title{Zero-shot Generative Large Language Models for Systematic Review Screening Automation}
		
	%
	%

\author{
	Shuai Wang\thanks{The University of Queensland, Australia.} \and
	Harrisen Scells\thanks{Leipzig University, Germany. } \and
	Shengyao Zhuang\thanks{CSIRO, Australia.} \and
	Martin Potthast\footnotemark[2] \thanks{Leipzig University and ScaDS.AI, Germany.} \and
	Bevan Koopman\footnotemark[3] \and
	Guido Zuccon\footnotemark[1]
}

\maketitle              

\begin{abstract}
	
Systematic reviews are crucial for evidence-based medicine as they comprehensively analyse published research findings on specific questions. Conducting such reviews is often resource- and time-intensive, especially in the screening phase, where abstracts of publications are assessed for inclusion in a review. This study investigates the effectiveness of using zero-shot large language models~(LLMs) for automatic screening. We evaluate the effectiveness of eight different LLMs and investigate a calibration technique that uses a predefined recall threshold to determine whether a publication should be included in a systematic review. 
Our comprehensive evaluation using five standard test collections shows that instruction fine-tuning plays an important role in screening, that calibration renders LLMs practical for achieving a targeted recall, and that combining both with an ensemble of zero-shot models saves significant screening time compared to state-of-the-art approaches.

\end{abstract}

\section{Introduction}

Systematic reviews are used extensively in medicine to comprehensively summarise all research findings on a specific question. Systematic reviews ensure a high level of rigour by including all and only those publications that meet predefined criteria, called the set of `included documents'.%
\footnote{Other commonly used terms are `studies', `research publications', and `references'.}
The selection of included documents starts with searching relevant databases such as PubMed~\cite{white2020pubmed} and the Cochrane Library~\cite{collaboration2002cochrane}. This search returns a list of `candidate documents', which are then screened for relevance and quality using the researchers' explicit inclusion and exclusion criteria.

Systematic reviews are labour-intensive and time-consuming, with most resources being invested in screening candidate documents, a process that can take months. While there are various methods to assist in optimizing the creation of systematic reviews (Section~\ref{related-work}),
one particular line of work focuses on minimising the number of documents that need to be manually screened. This has previously been pursued with classifiers to filter out documents that are not relevant, which may include manually labelling a significant number of the candidate documents to tune the classifier to the screening task at hand. Meanwhile, instruction-based generative large language models~(LLMs), such as OpenAI's ChatGPT,
\footnote{\url{https://chat.openai.com/}}
Llama~\cite{touvron2023llama}, and Alpaca~\cite{alpaca}, have demonstrated a remarkable ability to generate high-quality results in response to user instructions that often do not require task-specific tuning~\cite{alpaca,zhang2023generation}. In automating systematic reviews, these models have been fine-tuned for query formulation~\cite{ws2023chatgpt,wang2023generating}, as well as document classification and ranking~\cite{wang2023generating,robinson2023bio,alshami2023harnessing,EngeneAssessing2023}.

In this paper, we focus specifically on the use of \textit{zero-shot} large language models for the automatic screening of documents in systematic reviews (Section~\ref{method}). By `zero-shot', we mean using generative LLMs without explicitly optimising them for the screening task, which has the potential to relieve medical experts of any additional labelling burden. We examine two settings of our approach, an \textit{uncalibrated} and a \textit{calibrated} one. Both approaches prompt the model and use the probability of the next predicted (target) tokens to categorise documents as either `included' or `excluded'; the former directly uses the token with higher probability between `yes' and `no', the latter introduces the hyperparameter~$\theta$ as a new decision boundary of the classifier, calculated from the difference of the two tokens instead; $\theta$~is adjusted based on starting documents or previous systematic reviews.

In our evaluation, we address four research questions to investigate the factors that influence the effectiveness of the proposed zero-shot generative LLM-based automated screening method for systematic reviews (Section~\ref{evaluation-setup}):

\newcommand{\rqi}{How does the architecture and size of the LLMs influence effectiveness?}
\newcommand{\rqii}{How does instru\-ction-based fine-tuning influence effectiveness?}
\newcommand{\rqiii}{How does the calibration of the classifier's decisions with respect to the target tokens' likelihoods influence effectiveness?}
\newcommand{\rqiv}{How does ensembling LLM-based classifiers and current strong neural baselines influence effectiveness?} 

\begin{description}
\item[RQ1\label{rqi}] \rqi
\item[RQ2\label{rqii}] \rqii
\item[RQ3\label{rqiii}] \rqiii
\item[RQ4\label{rqiv}] \rqiv
\end{description}

Our evaluation results (Section~\ref{evaluation-results}) show that LlaMa2-7b-ins is currently the best model for this task, much better than the 13b~parameter variant. In general, instruction-based fine-tuning always outperforms the base models that have not been fine-tuned, and models based on LlaMa2 consistently outperform the baseline BERT-based method. Our approach also slightly outperforms (i.e. is competitive with) the fine-tuned Bio-SIEVE baseline. The calibrated setting of our method with ensembling achieves the best result overall and approaches the predefined recall target for the test topics, which indicates practical use.




\section{Related Work}
\label{related-work}

It is a requirement for high-quality systematic reviews to retrieve literature using a Boolean query~\cite{chandler2019cochrane,suhail2013methods}; the set of all retrieved documents must then be fully screened (assessed) for inclusion in the systematic review~\cite{chandler2019cochrane}. Research has explored the automatic creation of effective Boolean queries~\cite{scells2020comparison,scells2020conceptual,scells2020objective,scells2019refinement,ws2023chatgpt} (also with respect to the use of controlled vocabularies such as MeSH~\cite{wang2022automated,wang2021mesh,wang2023mesh}), and the ranking of the set of documents retrieved by the Boolean query (a task called ``screening prioritisation'')~\cite{miwa2014reducing,chen2017ecnu,alharbi2017ranking,wu2018ecnu,alharbi2018retrieving,lee2018seed,scells2017ltr,lee2018seed,lagopoulos2018learning,abualsaud2018system,zou2018technology,scells2020clf}, in order to begin downstream processes of the systematic review earlier~\cite{norman2019measuring}, e.g., acquiring the full-text of studies or results extraction.
The datasets that we consider in our experiments, including the CLEF TAR datasets~\cite{kanoulas2017clef,kanoulas2018clef,kanoulas2019clef}, specifically considered the task of screening prioritisation. In our paper, we consider a different task, the one of automating the screening phase of the systematic review; we discuss previous work related to this direction next.

Popular methods for automating the document screening phase are based on text classification~\cite{thomas2008methods}: a classifier is learned for an individual systematic review, typically in a supervised manner using labels obtained on a subset of the documents to be screened. Methods include traditional machine learning models like SVM~\cite{wallace2010semi,cohen2010prospective}, as well as classifiers based on encoder-based LLMs like BERT/BioBERT~\cite{robinson2023bio,aum2021srbert,carvallo2020neural}. Text classification methods are typically trained incrementally (acquiring labels through cycles of automatic classification) and often using active learning~\cite{wallace2012deploying,carvallo2020automatic,di2018interactive,di2017interactive,minas2018aristotle,anagnostou2017combining,singh2017iiit,yang2022goldilocks}. It is important to note that all of the methods above requires fine-tuning using labelled data specific to systematic review text classification in order to be effective.

	
In our work, we take a step further by considering the latest developments in generative LLMs to enhance the screening process. At the same time of developing this work, others have also explored similar directions. \cite{EngeneAssessing2023} employed ChatGPT for document screening, finding that ChatGPT's effectiveness is poor if the set of documents to be screened is imbalanced -- which is often the case in systematic reviews (i.e., typically, there are many more excluded documents than included among those that have been screened). Higher classification accuracy than ChatGPT was displayed by Bio-SIEVE~\cite{robinson2023bio}, a model fine-tuned from the Guanaco checkpoint~\cite{dettmers2023qlora}-- which in turn is based on the  Llama architecture. However, Bio-SIEVE also displayed severe consistency issues across review topics.
Importantly, both these works have notable limitations. 
The first study~\cite{EngeneAssessing2023} focused solely on the closed-sourced ChatGPT model. In addition, the evaluation was limited to only five systematic review topics and did not consider publicly available datasets with a broader range of review topics used in previous work. 
The second study~\cite{robinson2023bio} required to fine-tune the LLM, and relied on a self-constructed dataset for evaluation\footnote{Although the dataset is described to be public, it currently only contains the DOIs of the systematic review topics but not the labels, making reproduction difficult.}, limiting comparison with previous work. Furthermore, it only reported evaluation with respect to precision, recall, and accuracy; thus: (i) there is no account for the effect of class imbalance, (ii) there is no account that high-recall is considered essential in practice when conducting a systematic review. Conversely, in our work, we (1) consider open-sourced LLMs in a zero-shot setup, where further fine-tuning is not required, (2) take into account class imbalance and the high-recall nature of the task when evaluating methods, (3) rely on publicly available datasets that have been extensively used in previous work, thus facilitating comparison and reproduction.

\section{Generative LLMs for Automatic Document Screening}
\label{method}

\begin{figure*}[t!]
	\centering
	\includegraphics[width=0.97\textwidth]{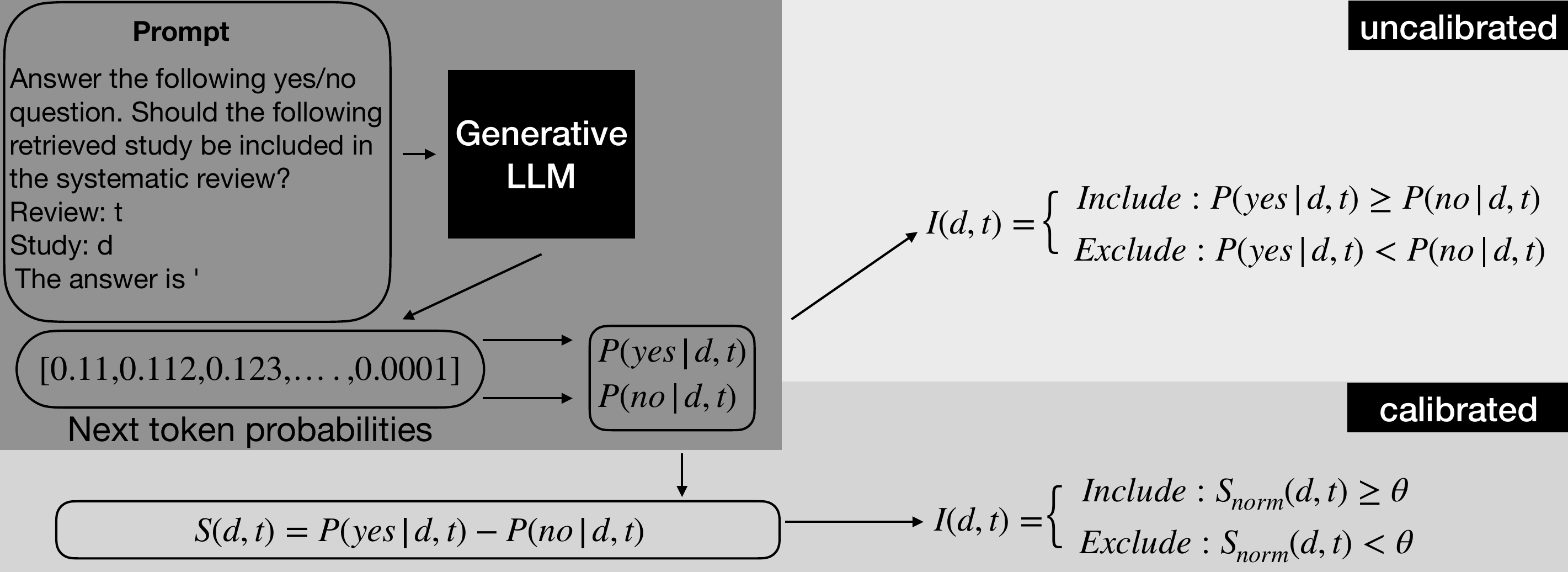}
	\caption{Our framework for automatic document screening using generative LLMs. $P(\texttt{yes}|d,t)$ ($ P(\texttt{no}|d,t)$) is the likelihood of the \texttt{yes} (\texttt{no}) token in the next token probability list, and $\theta$ is the decision boundary(threshold) used by the calibrated setting.}
	\label{fig:architecture}
\end{figure*}

Our framework for using a generative LLM for automatic document screening is shown in \autoref{fig:architecture}. The LLM considers a candidate document $d\in D$ for the systematic review topic $t\in T$; document screening is modelled as a classification task, using the function $I(d,t):D,T\rightarrow\{0,1\}$. Document $d$ is included for systematic review $t$ when $I(d,t)$ is $1$, and otherwise excluded. The function $I(d,t)$ is computed with respect to the output of the LLM for the prompt containing $d$ and $t$.
We investigate two instantiations of $I(d,t)$, uncalibrated and calibrated, which we explain below.


\par{\textbf{Uncalibrated Screening.}}
To determine whether a document should be included or not, uncalibrated screening directly compares the absolute values of the token likelihoods $P(\texttt{yes}|d,t)$ and $P(\texttt{no}|d,t)$ as generated by the LLM: 
\begin{equation*}
I(d,t) = 
\begin{cases}
1, & \text{if } P(\texttt{yes}|d,t) \geq P(\texttt{no}|d,t)\\
0, & \text{otherwise.}
\end{cases}
\end{equation*}

To ensure deterministic output, we forgo actual text generation with LLM. Instead, we represent the model decision using solely the probability of the next predicted token either to be `yes' or `no'. 
In this setting, the LLM returns an answer to the provided prompt of the decision with respect to the highest likelihood from the two tokens.

\par{\textbf{Calibrated Screening.}}
\label{sec:non-binary}
Building upon our uncalibrated instantiation, we calculate the difference between the likelihood of the next token to be \texttt{yes}, or \texttt{no}; then, we use a threshold to determine the inclusion of the document. We begin by computing the score $S(d,t)$ as the difference between the \texttt{yes} and \texttt{no} token likelihoods:

\begin{equation*}
	S(d,t) = 
	\begin{cases}
		P(\texttt{yes}|d,t) - P(\texttt{no}|d,t), &\text{if } P(\texttt{yes}|d,t) \geq P(\texttt{no}|d,t)\\
		0, & \text{otherwise.}
	\end{cases}
\end{equation*}

	
However, the probability distribution of the tokens depends on the individual documents, and thus is different across the documents. We then use min-max normalisation to normalise scores across all documents for a review topic $t$:
\begin{equation*}
	S_{\text{norm}}(d,t) = \frac{S(d,t) - \text{Min}(\{\forall d_i \in D : S(d_i,t)\}))}{\text{Max}(\{\forall d_i \in D : S(d_i,t)\}) - \text{Min}(\{\forall d_i \in D : S(d_i,t)\})}
\end{equation*}

Next, we identify a threshold $\theta$ using training data; $\theta$  is determined such that when used as the lower bound on scores for inclusion decisions, it ensures a minimum recall rate $k$.
Finally, we use $\theta$ to decide if a candidate document should be included: 

\begin{equation*}
I(d, t) = 
\begin{cases}
	1, & \text{if } S_{\text{norm}}(d,t) \ge \theta\\
	0, & \text{otherwise.} 
\end{cases}
\end{equation*}

The intuition behind exploring a calibrating screening approach is twofold. First, in the context of systematic review document screening, recall is of param-ount importance. For automation techniques to be used in practice, they must ensure the identification of all (or most) of the documents that should be included in the review. This is crucial because failing to capture all relevant documents may compromise the integrity of the review's conclusions and miss the main objective of a systematic review, that is its comprehensiveness. However, this focus on recall may not be naturally accounted for by LLMs, especially when accuracy is used to train/fine-tune classification models in the presence of highly imbalanced classes. 
Second, the inherent biases in different LLMs can lead to varying outcomes; some models may be naturally more inclusive, capturing a broader array of documents, while others may be more exclusive, being overly selective in their output. To account for these biases and to allow for customisation based on specific review needs, the calibrated instantiation of $I(d,t)$ offers a more adaptable and nuanced approach.

\par{\textbf{Ensembling of Screening Methods.}} We also consider an ensemble of screening methods. In particular, in our experiments we will ensemble the two most zero-shot effective LLMs and the BERT-based method we use as a comparative baseline. 
We use CombSUM to fuse the individual methods' decisions~\cite{kozorovitsky2011identical}. For Uncalibrated Screening, we directly combine the likelihoods of the model outputs. The decision rule \( I(d,t) \) is formulated as follows:
\begin{equation*}
I(d,t) = 
\begin{cases}
	1, & \text{if } \sum_{m \in \text{Methods}} P_{m}(\text{yes} | d, t) \geq \sum_{m \in \text{Methods}} P_{m}(\text{no} | d, t)\\
	0, & \text{otherwise.}
\end{cases}
\end{equation*}
For Calibrated Screening, we normalize \( S_{\text{norm}} \) to make the decisions:
\begin{equation*}
I(d, t) = 
\begin{cases}
	1, & \text{if } \sum_{m \in \text{Methods}}S_{\text{norm}}(d,t) \ge \theta\\
	0, & \text{otherwise.} 
\end{cases}
\end{equation*}

\section{Experimental Setup}
\label{evaluation-setup}

\subsection{Considered LLMs}

We employ an array of zero-shot generative LLMs that differ in architecture, training steps, and size (model parameters) to extensively evaluate their effectiveness for automatic systematic review document screening. 

	\textbf{LlaMa:} 
	The LlaMa series offers an open-sourced suite of decoder models with parameter sizes ranging from 7B to 65B. Exceptional in its zero-shot capabilities, LlaMa outperforms GPT-3 across multiple NLP benchmarks. These models leverage a rich and diverse training dataset of approximately 1.4 trillion tokens, harvested from various sources including web pages, code repositories, and Wikipedia~\cite{touvron2023llama}. 
	
	\textbf{Alpaca:} 
	Alpaca has been fine-tuned on the 7B-parameter LlaMa model according to the self-instruct methodology~\cite{wang2022self}.
	Alpaca's training corpus originates from the text-davinci-003 model~\footnote{\url{https://platform.openai.com/docs/models/gpt-3-5}}, initialized with 175 unique tasks.
	Preliminary assessments suggest that Alpaca, through instruction-based fine-tuning, achieves similar effectiveness to the OpenAI’s text-davinci-003 model~\cite{alpaca}.
	
	\textbf{Guanaco: } The Guanaco models stem from the LlaMa base models and are obtained through the memory-efficient 4-bit QLoRA fine-tuning on the OASST1 dataset~\cite{kopf2023openassistant,xu2023qa}. This represents a different fine-tuning strategy than that used in the other considered LLMs.
	Guanaco models have demonstrated competitive performance against commercial systems on the Vicuna and OpenAssistant benchmarks~\cite{chiang2023vicuna,kopf2023openassistant}.
	
	\textbf{Falcon:} 
	Falcon is available in two variants: Falcon-7B and Falcon-40B. These models were trained on large-scale corpora of 1 and 1.5 trillion tokens, respectively, primarily sourced from the RefinedWeb dataset~\cite{penedo2023refinedweb}. Notably, the Falcon family includes specialized ``instruct" versions — Falcon-7B-Instruct and Falcon-40B-Instruct — that excel in assistant-style tasks through fine-tuning on instructional and conversational datasets. 
	
	\textbf{LlaMa2:} 
	LlaMa2 extends the original LlaMa family, and comes in three parameter sizes: 7B, 13B, and 70B. Despite maintaining architectural similarity with its predecessor, LlaMa2 is trained on an expanded dataset of 2 trillion tokens, a 40\% increase from LlaMa~\cite{touvron2023llama,touvron2023llama2}. LlaMa2 also includes a specialized ``Chat" variant, LlaMa2 Chat, which incorporates advanced fine-tuning techniques such as ``Ghost Attention" for multi-turn dialogue consistency and an array of reinforcement learning methods~\cite{touvron2023llama2}.
	

Overall, we select eight models in our study:  LlaMa-7b,  Alpaca-7b-ins, Guana\-co-7b-ins, LlaMa2-7b, LlaMa2-13b, Falcon-7b-ins, LlaMa2-7b-ins, LlaMa2-13b-ins.%
\footnote{Note that for consistency of the paper, we name all instruction-tuned models with \textit{-ins}; The original names are: Alpaca-7b-ins: alpaca; Guanaco-7b-ins: guanaco-7b; Falcon-7b-ins: falcon-7b-instruct; LlaMa2-7b-ins: LlaMa2-7b-chat; LlaMa2-13b-ins: LlaMa2-13b-chat;} Table~\ref{table:prompts} demonstrates the prompts used for LLM-based automatic screening. Note that we do not include special tokens in the prompt due to page limit, specific prompt for each model are adapted based on their special token setup.
While we could have considered other models like the popular ChatGPT, their use can be financially prohibitive for our task. The predicted cost will be USD\$4,000 and USD\$80,000 if we use GPT-3.5-turbo and GPT-4, respectively. In our experiments, all employed models were configured to have a maximum token limit of 2048. This adjustment was particularly applied to the Alpaca model (original model has a limit of 512) and LlaMa2 models(original model has a limit of 4096) to ensure uniformity across all models. Consequently, we observed no instances of truncation in the experimental data.

\begin{table}[t!]
	\centering
	\scriptsize
	\caption{Input types and prompts designed for each model. Italicised text indicates values that are replaced with respective content.}
	\label{table:input-and-prompts-for-models}
	\begin{tabular}{c|M{320pt}}
\toprule
Model & \textnormal{Prompt} \\
\hline
\rotatebox[origin=c]{90}{Alpaca} & Below is an instruction that describes a task, paired with an input that provides further context. Write a response that appropriately completes the request. \linebreak \#\#\# Instruction:\linebreak 
Answer `yes' or `no'\ to Judge if the following retrieved study should be included by the systematic review? 
\linebreak \#\#\# Input: 
\linebreak Review: ${review\_title}$\linebreak Study: ${candidate\_document}$ \linebreak
\#\#\# Response: \\\hline

\rotatebox[origin=c]{90}{\parbox{1cm}{\centering All Other Models}} & 
Answer `yes' or `no'\ to Judge if the following retrieved study should be included by the systematic review? 
\linebreak Review: ${review\_title}$\linebreak Study: ${candidate\_document}$ \linebreak
The answer is $`$\\
\bottomrule

\end{tabular}
	\label{table:prompts}
\end{table}

\subsection{Datasets}

We experiment on the  CLEF TAR datasets and the Seed Collection dataset.
Four datasets were released as part of CLEF TAR~\cite{kanoulas2017clef,kanoulas2018clef,kanoulas2019clef}, covering different types of systematic reviews. The 2017 dataset contains 50  Diagnostic Test Accuracy (DTA) topics; 2018 adds 30 more; while in 2019, a dataset consisted of 8 DTA topics, while another included 40 intervention review (Int) topics. These datasets contain relevance assessments for about 600,000 documents in total, and for each topic, the review title and the protocol file are also provided. These datasets are distributed with standard train-test splits; however, because we consider the zero-shot capabilities of the investigated models, we do not use these splits and instead test on all available topics. 


The Seed Collection dataset consists of 39 review topics and over 50,000 candidate documents~\cite{wang2022little}.%
\footnote{We removed topic 18 as no relevant document exited in the candidate document list (the topic only contains one relevant document)}
 For each topic, the review title and inclusion/inclusion labels are provided along with a set of ``seed documents'': documents that were provided to the researcher designing the search strategy (query) for the review and that provide examples of documents related to the review (most are likely to meet the inclusion criteria, but it is possible some do not). In our experiment, we only evaluate based on the retrieved documents; included documents that are not in retrieved document set are removed.

\begin{table}[t!]
	\centering
	\tiny
	\caption{Comparison of uncalibrated results between baseline method and generative large language models. Statistical significance, determined by a Student’s two-tailed paired t-test with Bonferroni correction ($p < 0.05$), between the top-performing method \textit{LlaMa2-7b-ins} and others is marked by *.}
	\begin{minipage}[t]{\textwidth}
	\centering
\begin{adjustbox}{valign=t}
\begin{tabular}{p{2pt}l|llllll}
	\toprule
	& Model &P & R & B-AC&F3&Suc&WSS \\
	\hline \addlinespace[2pt]
\multirow{9}{*}{\rotatebox{90}{CLEF-2017}} & BioBERT & 0.06 & 0.95* & 0.61* & 0.30 & 0.74* & 0.26* \\\cmidrule{2-8}
& LlaMa-7b & 0.04* & 0.92* & 0.48* & 0.24* & 0.46* & 0.03* \\
& LlaMa2-7b & 0.07 & 0.50* & 0.60* & 0.23* & 0.02* & 0.70* \\
& LlaMa2-13b & 0.04* & 1.00* & 0.50* & 0.25* & 0.98* & 0.00* \\
& Falcon-7b-ins & 0.05* & 0.92* & 0.52* & 0.25* & 0.44 & 0.12* \\
& Alpaca-7b-ins & 0.04* & 0.92* & 0.51* & 0.25* & 0.38 & 0.11* \\
& LlaMa2-7b-ins & 0.08 & 0.87 & \textbf{0.72} & \textbf{0.35} & 0.26 & 0.56 \\
& LlaMa2-13b-ins & \textbf{0.19*} & 0.41* & 0.66* & 0.31 & 0.04* & \textbf{0.91*} \\
& Guanaco-7b-ins & 0.04* & \textbf{1.00*} & 0.50* & 0.25* & \textbf{1.00*} & 0.00* \\\midrule

\multirow{9}{*}{\rotatebox{90}{CLEF-2018}} & BioBERT & 0.06 & 0.97* & 0.59* & 0.9 & 0.87* & 0.19* \\\cmidrule{2-8}
& LlaMa-7b & 0.05* & 0.92* & 0.48* & 0.25* & 0.33 & 0.04* \\
& LlaMa2-7b & 0.07 & 0.49* & 0.59* & 0.22* & 0.03* & 0.69* \\
& LlaMa2-13b & 0.05* & 1.00* & 0.50* & 0.26 & \textbf{1.00*} & 0.00* \\
& Falcon-7b-ins & 0.05* & 0.92 & 0.51* & 0.25* & 0.40 & 0.11* \\
& Alpaca-7b-ins & 0.05* & 0.91 & 0.51* & 0.25* & 0.30 & 0.11* \\
& LlaMa2-7b-ins & 0.09 & 0.88 & \textbf{0.75} & \textbf{0.37} & 0.27 & 0.59 \\
& LlaMa2-13b-ins & \textbf{0.26*} & 0.36* & 0.66* & 0.30 & 0.00* & \textbf{0.94*} \\
& Guanaco-7b-ins & 0.05* & \textbf{1.00*} & 0.50* & 0.26 & \textbf{1.00*} & 0.00* \\\midrule
\multirow{9}{*}{\rotatebox{90}{CLEF-2019-dta}} & BioBERT & 0.07 & 0.99 & 0.58 & 0.30 & 0.88 & 0.18* \\\cmidrule{2-8}
& LlaMa-7b & 0.07 & 0.93 & 0.48* & 0.27 & 0.25 & 0.03* \\
& LlaMa2-7b & 0.08 & 0.48* & 0.58* & 0.23 & 0.00* & 0.68 \\
& LlaMa2-13b & 0.07 & 1.00 & 0.50* & 0.28 & \textbf{1.00} & 0.00* \\
& Falcon-7b-ins & 0.07 & 0.95 & 0.54* & 0.29 & 0.50 & 0.12* \\
& Alpaca-7b-ins & 0.07 & 0.91 & 0.52* & 0.28 & 0.25 & 0.12* \\
& LlaMa2-7b-ins & 0.09 & 0.92 & \textbf{0.71} & \textbf{0.35} & 0.62 & 0.49 \\
& LlaMa2-13b-ins & \textbf{0.19} & 0.49* & 0.69 & 0.32 & 0.00* & \textbf{0.87*} \\
& Guanaco-7b-ins & 0.07 & \textbf{1.00} & 0.50* & 0.28 & \textbf{1.00} & 0.00* \\\midrule
\end{tabular}
\end{adjustbox}
\end{minipage}
\hfill
\begin{minipage}[t]{\textwidth}
	\centering
\begin{adjustbox}{valign=t}
\begin{tabular}{p{2pt}l|llllll}
	\toprule
	& Model &P & R& B-AC&F3&Suc&WSS \\
	\hline \addlinespace[2pt]
	\multirow{9}{*}{\rotatebox{90}{CLEF-2019-Int}} & BioBERT & 0.10 & 0.98* & 0.58* & 0.32 & 0.90* & 0.16* \\\cmidrule{2-8}
	& LlaMa-7b & 0.05* & 0.86 & 0.47* & 0.26 & 0.30 & 0.08* \\
	& LlaMa2-7b & 0.08 & 0.30* & 0.55* & 0.18* & 0.05* & 0.80* \\
	& LlaMa2-13b & 0.05 & 1.00* & 0.50* & 0.29 & 0.97* & 0.00* \\
	& Falcon-7b-ins & 0.05 & 0.91 & 0.50* & 0.27 & 0.57 & 0.09* \\
	& Alpaca-7b-ins & 0.05 & 0.87 & 0.49* & 0.27 & 0.30 & 0.12* \\
	& LlaMa2-7b-ins & 0.08 & 0.90 & \textbf{0.70} & \textbf{0.35} & 0.42 & 0.48 \\
	& LlaMa2-13b-ins & \textbf{0.17*} & 0.45* & 0.67 & 0.33 & 0.05* & \textbf{0.87*} \\
	& Guanaco-7b-ins & 0.05 & \textbf{1.00*} & 0.50* & 0.29 & \textbf{1.00*} & 0.00* \\\midrule
	\multirow{9}{*}{\rotatebox{90}{Seed Collection}} & BioBERT & 0.04 & 0.93 & 0.54* & 0.24 & 0.77* & 0.16* \\\cmidrule{2-8}
	& LlaMa-7b & 0.04 & 0.89 & 0.48* & 0.21 & 0.56 & 0.07* \\
	& LlaMa2-7b & 0.04 & 0.29* & 0.53* & 0.15* & 0.03* & 0.78* \\
	& LlaMa2-13b & 0.04 & \textbf{1.00*} & 0.50* & 0.23 & \textbf{1.00*} & 0.00* \\
	& Falcon-7b-ins & 0.04 & 0.93 & 0.50* & 0.22 & 0.69 & 0.07* \\
	& Alpaca-7b-ins & 0.04 & 0.90 & 0.50* & 0.22 & 0.49 & 0.10* \\
	& LlaMa2-7b-ins & 0.05 & 0.90 & 0.66 & 0.27 & 0.54 & 0.40 \\
	& LlaMa2-13b-ins & \textbf{0.13*} & 0.48* & \textbf{0.67} & \textbf{0.28} & 0.05* & \textbf{0.85*} \\
	& Guanaco-7b-ins & 0.04 & \textbf{1.00*} & 0.50* & 0.23 & \textbf{1.00*} & 0.00* \\\midrule
\end{tabular}
\end{adjustbox}
\end{minipage}

	\label{table:review-title-uncalibrated}
\end{table}

\section{Results}
\label{evaluation-results}


\subsection{Baseline}



We compare the effectiveness of zero-shot LLMs against a baseline that relies on the BERT architecture but uses a domain-specific variant as backbone: BioBERT~\cite{lee2020biobert,wang2022neural}. BioBERT employs the same architecture as BERT, but the corpus used for self-supervised training contains biomedical text (instead of general domain text like for BERT). BioBERT has been shown effective across a range of applications related to health tasks, including for screening prioritisation on medical systematic reviews on the datasets we consider~\cite{wang2022neural}, and thus is a strong baseline.
To use BioBERT in our text classification task, we concatenate the topic title with the candidate document to form the input to the backbone. A classification head based on a sigmoid activation function is then used to determine the inclusion of a candidate document for the specified topic.%
\footnote{In the uncalibrated setting for BioBERT, we established a decision threshold of 0.5 to determine the inclusion of a document in a review topic. Specifically, a document is included if the BioBERT output satisfies the condition $output \geq 0.5$; otherwise, it is excluded.}


\begin{table}[t!]
	\centering
	\tiny
	\caption{Comparison between the Calibrated (Cal) and Uncalibrated (Unc) approaches using the BioBERT model, LlaMa2-7b-ins model (7b-ins), the LlaMa2-13b-ins model (13b-ins) and the Ensemble of the three models (Ensemb). The calibrated method's number or character in the bracket () denotes the pre-set target recall (0.95  \& 1) or using seed documents (S). Statistical significance for each generative model across different datasets is assessed using a Student's two-tailed paired t-test with a Bonferroni correction ($p < 0.05$) with respect to the uncalibrated approach, denoted by *. The highest evaluated scores for \textit{each dataset} are bolded.}
	\begin{minipage}[t]{\textwidth}
	\centering
	\begin{adjustbox}{valign=t}
		
		\begin{tabular}{p{2pt}ll|llllll}
	\toprule
	& Model & Setting &P & R& B-AC&F3&Suc&WSS \\
	\hline \addlinespace[2pt]
\multirow{15}{*}{\rotatebox{90}{CLEF-2017}} & \multirow{2}{*}{\rotatebox{45}{\shortstack{Bio\\BERT}}} & Unc & 0.06 & 0.95 & 0.61 & 0.30 & 0.74 & 0.26  \\
& & Cal(0.95)  & 0.06 & 0.92 & 0.64 & 0.31 & 0.50* & 0.34*\\
& & Cal(1) & 0.06 & 0.97 & 0.60 & 0.29 & 0.82 & 0.23 \\

\cmidrule{2-9}
& \multirow{3}{*}{\rotatebox{45}{7b-ins}} & Unc & 0.08 & 0.87 & 0.72 & 0.35 & 0.26 & 0.56\\
& & Cal(0.95) & 0.06* & 0.92* & 0.69* & 0.32 & 0.52 & 0.44 \\
& & Cal(1) & 0.05* & \textbf{0.99}* & 0.60* & 0.28 & 0.96 & 0.20 \\
\cmidrule{2-9}
& \multirow{3}{*}{\rotatebox{45}{13b-ins}} & Unc & 0.19 & 0.41 & 0.66 & 0.31 & 0.04 & 0.91\\
& & Cal(0.95) & 0.06* & 0.93 & 0.59* & 0.28 & 0.50* & 0.25* \\
& & Cal(1) & 0.05* & 0.98 & 0.53* & 0.26 & 0.88* & 0.08* \\
\cmidrule{2-9}
& \multirow{3}{*}{\rotatebox{45}{Ensemb}} & Unc & \textbf{0.31} & 0.13 & 0.56 & 0.13 & 0.00 & \textbf{0.98}\\
& & Cal(0.95) & 0.08 & 0.93* & \textbf{0.72} & \textbf{0.35}* & 0.52* & 0.50* \\
& & Cal(1) & 0.06 & 0.97* & 0.63 & 0.30 & \textbf{0.90}* & 0.29* \\
\midrule

\multirow{15}{*}{\rotatebox{90}{CLEF-2018}} & \multirow{2}{*}{\rotatebox{45}{\shortstack{Bio\\BERT}}} & Unc & 0.06 & 0.97 & 0.59 & 0.29 & 0.87 & 0.19  \\
& & Cal(0.95)  & 0.07 & 0.91* & 0.63 & 0.30 & 0.57* & 0.33*\\
& & Cal(1) & 0.06 & 0.97 & 0.59 & 0.29 & 0.87 & 0.21 \\
\cmidrule{2-9}
& \multirow{3}{*}{\rotatebox{45}{7b-ins}} & Unc & 0.09 & 0.88 & 0.75 & 0.37 & 0.27 & 0.59\\
& & Cal(0.95) & 0.08* & 0.94* & 0.71* & 0.35* & 0.50 & 0.46 \\
& & Cal(1) & 0.06* & \textbf{0.99}* & 0.62* & 0.30 & \textbf{1.00} & 0.24 \\
\cmidrule{2-9}
& \multirow{3}{*}{\rotatebox{45}{13b-ins}} & Unc & 0.26 & 0.36 & 0.66 & 0.30 & 0.00 & 0.94\\
& & Cal(0.95) & 0.06 & 0.94* & 0.59* & 0.29 & 0.47* & 0.22* \\
& & Cal(1) & 0.05 & 0.97 & 0.53* & 0.27 & 0.80* & 0.08* \\
\cmidrule{2-9}
& \multirow{3}{*}{\rotatebox{45}{Ensemb}} & Unc & \textbf{0.35} & 0.12 & 0.54 & 0.12 & 0.00 & \textbf{0.95}\\
& & Cal(0.95) & 0.09* & 0.94* & \textbf{0.75} & \textbf{0.38}* & 0.50* & 0.54* \\
& & Cal(1) & 0.06 & 0.99* & 0.64 & 0.32* & 0.93* & 0.28* \\
\midrule

\multirow{15}{*}{\rotatebox{90}{CLEF-2019-dta}}  & \multirow{2}{*}{\rotatebox{45}{\shortstack{Bio\\BERT}}} & Unc & 0.07 & 0.99 & 0.58 & 0.30 & 0.88 & 0.18 \\
& & Cal(0.95)   & 0.08 & 0.89 & 0.59 & 0.26 & 0.50 & 0.27\\
& & Cal(1) & 0.08 & 0.91 & 0.59 & 0.27 & 0.62 & 0.25 \\

\cmidrule{2-9}

& \multirow{3}{*}{\rotatebox{45}{7b-ins}} & Unc & 0.09 & 0.92 & 0.71 & \textbf{0.35} & 0.62 & 0.49\\
& & Cal(0.95) & 0.10* & 0.91* & 0.71* & 0.34 & 0.50 & 0.50 \\
& & Cal(1) & 0.08* & 0.97* & 0.66 & 0.32 & 0.75 & 0.34 \\
\cmidrule{2-9}
& \multirow{3}{*}{\rotatebox{45}{13b-ins}} & Unc & 0.19 & 0.49 & 0.69 & 0.32 & 0.00 & 0.87\\
& & Cal(0.95) & 0.08 & 0.95 & 0.56 & 0.29 & 0.50* & 0.16* \\
& & Cal(1) & 0.07 & 0.99 & 0.51 & 0.28 & 0.88* & 0.03* \\
\cmidrule{2-9}
& \multirow{3}{*}{\rotatebox{45}{Ensemb}} & Unc & \textbf{0.31} & 0.21 & 0.59 & 0.19 & 0.00 & \textbf{0.96}\\
& & Cal(0.95) & 0.10 & 0.91 & \textbf{0.73} & 0.34* & 0.50* & 0.52* \\
& & Cal(1) & 0.09* & \textbf{0.99} & 0.64 & 0.32* & \textbf{1.00}* & 0.28* \\
\midrule

\end{tabular}
\end{adjustbox}
\end{minipage}
\hfill
\begin{minipage}[t]{\textwidth}
	\centering
\begin{adjustbox}{valign=t}
\begin{tabular}{p{2pt}ll|llllll}
	\toprule
& Model & Setting &P & R& B-AC&F3&Suc&WSS \\
\hline \addlinespace[2pt]
\multirow{15}{*}{\rotatebox{90}{CLEF-2019-Int}} & \multirow{2}{*}{\rotatebox{45}{\shortstack{Bio\\BERT}}} & Unc & 0.10 & 0.98 & 0.58 & 0.32 & \textbf{0.90} & 0.16  \\
& & Cal(0.95)  & 0.10 & 0.87* & 0.59 & 0.29 & 0.50* & 0.31*\\
& & Cal(1) & 0.10 & 0.90* & 0.59 & 0.30 & 0.62* & 0.27* \\

\cmidrule{2-9}
& \multirow{3}{*}{\rotatebox{45}{7b-ins}} & Unc & 0.08 & 0.90 & 0.70 & 0.35 & 0.42 & 0.48\\
& & Cal(0.95) & 0.08* & 0.91* & 0.67* & 0.34 & 0.50 & 0.42 \\
& & Cal(1) & 0.07* & 0.93* & 0.64* & 0.33 & 0.65 & 0.34 \\
\cmidrule{2-9}
& \multirow{3}{*}{\rotatebox{45}{13b-ins}} & Unc & 0.17 & 0.45 & 0.67 & 0.33 & 0.05 & 0.87\\
& & Cal(0.95) & 0.07* & 0.90 & 0.58 & 0.30 & 0.50* & 0.25* \\
& & Cal(1) & 0.06* & 0.94 & 0.55 & 0.29 & 0.62* & 0.16* \\
\cmidrule{2-9}
& \multirow{3}{*}{\rotatebox{45}{Ensemb}} & Unc & \textbf{0.35} & 0.23 & 0.58 & 0.22 & 0.05 & \textbf{0.92}\\
& & Cal(0.95) & 0.09* & 0.93* & \textbf{0.70} & \textbf{0.37}* & 0.50* & 0.45* \\
& & Cal(1) & 0.08* &\textbf{0.96}* & 0.67 & 0.35* & 0.68* & 0.35* \\

\midrule
\multirow{19}{*}{\rotatebox{90}{Seed Collection}} & \multirow{4}{*}{\rotatebox{45}{\shortstack{Bio\\BERT}}} & Unc & 0.04 & 0.93 & 0.54 & 0.24 & 0.77 & 0.16\\
& & Cal(0.95)  & 0.05 & 0.80* & 0.55 & 0.22 & 0.50* & 0.29* \\
& & Cal(1) & 0.05 & 0.83 & 0.55 & 0.23 & 0.53* & 0.26 \\
& & Cal (S) & 0.04 & 0.93 & 0.54 & 0.23 &  0.76 & 0.15 \\

\cmidrule{2-9}
& \multirow{4}{*}{\rotatebox{45}{7b-ins}} & Unc & 0.05 & 0.90 & 0.66 & 0.27 & 0.54 & 0.40\\
& & Cal(0.95) & 0.05* & 0.90* & 0.66 & 0.28* & 0.51 & 0.41 \\
& & Cal(1) & 0.05* & 0.92* & 0.65 & 0.27* & 0.56 & 0.38 \\
 & & Cal (S) & 0.05 & 0.97* & 0.6* & 0.26 & 0.77* & 0.22* \\
\cmidrule{2-9}
& \multirow{4}{*}{\rotatebox{45}{13b-ins}} & Unc & 0.13 & 0.48 & 0.67 & 0.28 & 0.05 & 0.85\\
& & Cal(0.95) & 0.06* & 0.87 & 0.64* & 0.27* & 0.51* & 0.39* \\
& & Cal(1) & 0.05* & 0.93 & 0.59* & 0.26 & 0.59* & 0.25* \\
 & & Cal (S) & 0.06* & 0.87* & 0.63 & 0.29 & 0.54* & 0.38* \\
\cmidrule{2-9}
& \multirow{4}{*}{\rotatebox{45}{Ensemb}} & Unc & \textbf{0.16} & 0.18 & 0.52 & 0.14 & 0.00 & \textbf{0.86}\\
& & Cal(0.95) & 0.07* & 0.86* & \textbf{0.71} & 0.31* & 0.49* & 0.53* \\
& & Cal(1) & 0.07* & 0.88* & 0.70 & \textbf{0.30}* & 0.56* & 0.49* \\
& & Cal (S) & 0.04* & \textbf{1.00}* & 0.55 & 0.25* & \textbf{0.97}* & 0.10* \\
\midrule

\end{tabular}
\end{adjustbox}
\end{minipage}

	\label{table:review-title-calibrated}
\end{table}


\subsection{Evaluation Measures}

We use set-based metrics for evaluation: precision, recall, and F-3, which emphasize the importance of recall over precision. Additionally, we adopt balanced accuracy (B-AC) as a pivotal metric, as it particularly suits the nature of the systematic review document screening task, where excluded documents substantially outnumber included ones;
$
\text{B-AC} = \frac{1}{2} \left( \frac{\text{TP}}{\text{TP} + \text{FN}} + \frac{\text{TN}}{\text{TN} + \text{FP}} \right)
$.
We also report the success rate, which quantifies the fraction of topics achieving a pre-specified target recall. We adopt a representative target recall of 0.95, a standard threshold for systematic review document screening~\cite{bramer2017optimal,crumley2005resources,callaghan2020statistical}: often systems that do not achieve at least 0.95 recall are deemed of no practical use for systematic review automation.
Lastly, we gauge the efficiency of automatic document screening using the Work Saved by Sampling at a specific recall level (\(WSS\))~\cite{cohen2006reducing}. This is expressed as:
$
WSS = \frac{TN + FN}{N} - (1-r)
$
where \(N\) denotes the total sample count and \(r\) signifies the recall level; we set \(r\) to 1, representing  total recall.

\subsection{Threshold Setting}

For the calibration setting, the threshold $\theta$ value needs to be set. We devise two approaches to determine $\theta$:

\begin{enumerate}
\item
\textbf{Extrapolation from Collection}: we perform a leave-one-out experiment across all systematic review topics in a dataset. We identify threshold values that have consistently yielded robust results in the sample topics (all other topics except the target topic)—optimizing for a high recall rate—using the median score of candidate documents that achieved the target recall. The obtained threshold is then applied to the target topic under consideration. Note that cross-validation is used to determine the $\theta$ value only: the LLMs are still zero-shot. This is, however, a somewhat artificial setting, in that if training material was available for determining $\theta$, then it could also be used to tune the LLMs (though computational costs may prevent this but our method does not require training of the LLM itself). We will consider a more appropriate option next.
\item 
\textbf{Calibration with Seed Studies}: we employ the uncalibrated LLM to generate inclusion scores for a set of seed studies (exemplar documents that are often identified prior to searching and screening). If the lowest score for a seed study is below the classifier's threshold for inclusion (decision boundary), then the threshold is lowered to the score obtained by that seed study: we use this as the new threshold for the calibrated LLM. This adjustment aims to improve recall.
Typical targets for recall for systematic review are 0.95 or 1; we then experiment with these values to determine \( \theta \). 
\end{enumerate}

\begin{table}[t]
	\tiny
	\centering
	\caption{Comparison of Fine-tuned baseline to our method; Statistical significance, determined by a Student’s two-tailed paired t-test with Bonferroni correction ($p < 0.05$), between Uncalibrated \textit{Bio-SIEVE} method and others is marked by *. }
	\begin{tabular}{ll|llllll}
	\toprule
	Model & Setting&P& R &B-AC& F3 &Suc &WSS\\
	\hline \addlinespace[2pt]
\multirow{5}{*}{Bio-SIEVE} & Original/Calibrated& \textbf{0.232} & 0.576 & 0.727 & \textbf{0.429} & 0.111 & 0.858 \\\cmidrule{2-8} 
 & Calibrated(Recall=0.95) & 0.102* & 0.877* & 0.683 & 0.348 & 0.481* & 0.471* \\ \cmidrule{2-8}
&Calibrated (Recall=1) & 0.088* & 0.945* & 0.666 & 0.339 & 0.704* & 0.369* \\ \midrule

\multirow{5}{*}{LlaMa2-7b-ins}& Uncalibrated & 0.078* & 0.920* & 0.725 & 0.359 & 0.333 & 0.513* \\\cmidrule{2-8}
 & Calibrated (Recall=0.95) & 0.068* & 0.935* & 0.685 & 0.333 & 0.481* & 0.421* \\ \cmidrule{2-8}
 & Calibrated(Recall=1) & 0.059* & \textbf{0.990*} & 0.621* & 0.311 & \textbf{1.000}* & 0.241* \\ \midrule
%

\multirow{5}{*}{Ensemble} &Uncalibrated  & 0.400* & 0.204* & 0.594* & 0.199* & 0.037 & \textbf{0.972}* \\\cmidrule{2-8}
& Calibrated (Recall=0.95) & 0.095* & 0.937* & \textbf{0.729} & 0.373 & 0.519* & 0.500* \\ \cmidrule{2-8}
 & Calibrated (Recall=1) & 0.068* & 0.981* & 0.630* & 0.322 & 0.889* & 0.266* \\ \midrule
\end{tabular}
	\label{table:with-fine-tuned-baseline}
\end{table}

\par{\textbf{RQ1: Architecture and Size of Model.}}
Consider the results reported in Table~\ref{table:review-title-uncalibrated}.
For \textit{model architecture}, we compare four models: \textit{Falcon-7b-ins}, \textit{Alpaca-7b-ins}, \textit{LlaMa2-7b-ins} and \textit{Guanaco-2-7b-ins} --- all of which have the same number of parameters. The results indicate that \textit{LlaMa2-7b-ins} is the most effective for the task, outperforming the others across all evaluation metrics except recall and success rate. Specifically, this model obtained a high WSS while incurring only a marginal drop in recall: a significant loss was observed only on CLEF-2017. Concerning success rate, \textit{LlaMa2-7b-ins} exhibited comparable performance to its counterparts, showing no statistically significant differences.

For \textit{model size}, we consider two variants of the \textit{LlaMa2-ins} architecture: one with 7 billion parameters (\textit{LlaMa2-7b-ins}) and another with 13 billion parameters (\textit{LlaMa2-13b-ins}). Our findings suggest a trade-off between recall and WSS. Specifically, the 7-billion parameter variant obtains significantly higher recall, but this comes at the expense of reduced savings, evidenced by significantly lower WSS. Regarding B-AC, \textit{LlaMa2-7b-ins} generally outperforms its larger counterpart across multiple datasets, except for the Seed Collection. Statistically significant differences in B-AC were only noted for CLEF-2017 and CLEF-2018.

\par{\textbf{RQ2: Impact of instruction fine-tuning.}}
Consider again Table~\ref{table:review-title-uncalibrated}. We contrast instruction-fine-tuned models against their base counterparts: \textit{LlaMa2-7b-ins} VS. \textit{LlaMa2-7b}, \textit{LlaMa2-13b-ins} VS. \textit{LlaMa2-13b}, \textit{Alpaca-7b-ins} VS. \textit{LlaMa-7b}. Across all differences, a significant improvement in B-AC is observed. Nevertheless, the models exhibit divergent behaviours in other metrics. For \textit{LlaMa-7b} and \textit{LlaMa2-13b}, fine-tuning leads to higher WSS at the expense of reduced recall. Conversely, \textit{LlaMa2-7b-ins} exhibits a significant decline in WSS but obtains higher recall, success rate, and F3 except in the CLEF-2019-dta, where the F3 improvement is not statistically significant. We also conducted a comparative evaluation with \textit{Guanaco-7b-ins}, a QLoRA fine-tuned model. While it does outperform \textit{LlaMa-7b} in B-AC, the model classifies all candidate documents as relevant, nullifying any practical applicability for systematic review screening.

In summary, our analyses suggest that instruction-based fine-tuning is generally beneficial for improving document screening accuracy. However, the specific gains --- whether in savings or recall --- depend on the base model's architecture. Our experiments also suggest that QLoRa fine-tuning does not yield an effective model for this particular task.

\par{\textbf{RQ3: Impact of Calibration.}} Consider Table~\ref{table:review-title-calibrated}.
We find that calibrated models reliably meet their pre-set recall targets and provide an attractive solution for practical implementation for automatic document screening.
Specifically, in our tests that considered the extrapolation from collection calibration, approximately 50\% of the topics met the pre-set recall target of 0.95 by comparing success rates obtained in each dataset (note that success rate in our experiments is set to measure a 0.95 recall level). This further improves (success rate between 0.56 and 1.00) when the target recall for determining the threshold is set to 1.
We further compare the performance of three calibrated models, \textit{BioBERT}, \textit{LlaMa2-7b-ins} and \textit{LlaMa2-13b-ins}. Generally, the 7-billion parameter LlaMa2 model significantly outperforms the two other models in both B-AC and WSS. As for success rate and recall, the models exhibit similar effectiveness; \textit{LlaMa2-7b-ins} performs the same or better in 60\% of the cases for success rate and in 40\% of cases for average recall.

The calibration with seed documents method could only be tested on the Seed Collection, as CLEF datasets have no seed studies. In this case, \textit{LlaMa2-7b-ins} consistently obtains higher recall: 70\% of topics achieved perfect recall, compared to only 50\% using the other calibration method. Although calibration with seed studies generally improves recall, our analysis indicates that \textit{LlaMa2-13b-ins} displays more volatile effectiveness in this setting, possibly due to the varying quality and quantity of seed documents across different topics.

\par{\textbf{Ensemble of Automatic Screening Methods.}} Consider Table~\ref{table:review-title-calibrated} with respect to the Ensemble results, obtained by ensambling \textit{LlaMa2-7b-ins}, \textit{LlaMa2-13b-ins} and the \textit{BioBERT} baseline.
The Ensemble strategy yields consistently higher B-AC and WSS, when calibrated. Moreover, when pitted against individual generative LLMs calibrated with the same threshold recall, the Ensemble method obtains higher WSS, precision, and F3. Exceptions are observed in CLEF-2018 and Seed Collection, where the Ensemble strategy registers lower success rates. Interestingly, the Ensemble's performance dips in recall when not calibrated. This decline may be attributed to the model's aggressive document exclusion strategy, as evidenced by its consistently high WSS across datasets. Overall, our findings indicate that a calibrated Ensemble approach generally outperforms single generative LLMs.

%
%



\section{Discussion and Outlook}

\par{\textbf{Comparison with fine-tuned LLMs.}}
Although this study aimed to investigate the effectiveness of zero-shot generative LLMs in systematic review document screening, we are also interested in comparing our method to the state-of-the-art fine-tuned model. For this comparison, we consider the Bio-SIEVE approach, a fine-tuned model for systematic review document screening, and compare it with our best methods in Table~\ref{table:with-fine-tuned-baseline}.%
\footnote{Comparison is however not straightforward as Bio-SIEVE used most of the datasets we consider here for fine-tuning; we then evaluate effectiveness using the only 27 topics from CLEF-TAR that were not used to fine-tune Bio-SIEVE.}
We also apply our calibration approach to Bio-SIEVE. Surprisingly the most effective model, LlaMa2-7b-ins, obtains a B-AC comparable to Bio-SIEVE, and our Ensemble method is even more effective than Bio-SIEVE, although differences are not significant. 

Another noteworthy observation is Bio-SIEVE's low recall and success rate, especially when not calibrated (original). These results raise concerns regarding Bio-SIEVE's practical utility for the screening task, as a low recall is often not accepted by the researchers conducting the review as it translates into missing important studies.
While calibration improves Bio-SIEVE's recall, this is still inferior to our zero-shot model under the same calibration setting. This finding suggests that although fine-tuning can improve effectiveness, it requires careful calibration for systematic review document screening. Looking forward, fine-tuning remains an interesting avenue for research but may necessitate alternative calibration strategies for practical utility for this task.

\par{\textbf{Variation in model input prompt.}} 
While we only considered one type of prompt for each model, it is important to highlight that generative LLMs are sensitive to prompt formulation~\cite{yang2023large,zhao2021calibrate,lu2021fantastically}. 
Due to page constraints, we could not deeply discuss the effects of alternative prompt formulations, such as those based on inclusion/exclusion criteria or seed studies. However, preliminary investigations into these aspects show a similar trend to what is observed when solely using review topic titles as prompts. These additional results are provided in a supplementary digital appendix for completeness.%
\footnote{\url{https://github.com/ielab/ECIR-2024-llm-screening}}

\section{Conclusion}

We comprehensively evaluated zero-shot LLMs for systematic review document screening and introduced a calibration method for tuning the model output. We further explored the utility of an ensemble method that combines the top zero-shot LLMs with the BioBERT baseline.

Our results highlight the importance of output calibration when applying generative LLMs to systematic review document screening. This calibration maintains review quality and reliably by meeting pre-set recall targets, thus offering the flexibility to adjust the model to the specific requirements of a systematic review. Furthermore, when calibrated, our ensemble method outperforms the current state-of-the-art fine-tuned model, Bio-SIEVE~\cite{robinson2023bio}. 
We also emphasized the role of instruction-based fine-tuning in effectively leveraging generative LLMs for this application, while we showed that QLoRa-tuning does not yield effective results for this task. 

The findings reported in the paper suggest that LLM-based methods can be created for automatically screening documents for systematic reviews, leading to considerable savings in manual effort. Furthermore, this can be done without requiring expensive fine-tuning (both in terms of labelling and computation). The fact that a high recall level can be obtained across a large number of different types of reviews suggests that these methods might be mature enough for actual adoption in systematic review workflows.

\bibliographystyle{splncs04}
\bibliography{bibliography}

\begin{thebibliography}{10}
\providecommand{\url}[1]{\texttt{#1}}
\providecommand{\urlprefix}{URL }
\providecommand{\doi}[1]{https://doi.org/#1}

\bibitem{abualsaud2018system}
Abualsaud, M., Ghelani, N., Zhang, H., Smucker, M.D., Cormack, G.V., Grossman,
  M.R.: A system for efficient high-recall retrieval. In: Proceedings of the
  41st Annual International {{ACM SIGIR}} Conference on Research and
  Development in Information Retrieval. pp. 1317--1320 (2018)

\bibitem{alharbi2018retrieving}
Alharbi, A., Briggs, W., Stevenson, M.: Retrieving and ranking studies for
  systematic reviews: {{University}} of {{Sheffield}}'s approach to {{CLEF
  eHealth}} 2018 {{Task}} 2. In: {{CEUR}} Workshop Proceedings: {{Working}}
  Notes of {{CLEF}} 2018: {{Conference}} and Labs of the Evaluation Forum.
  vol.~2125. {CEUR Workshop Proceedings} (2018)

\bibitem{alharbi2017ranking}
Alharbi, A., Stevenson, M.: Ranking abstracts to identify relevant evidence for
  systematic reviews: {{The}} university of sheffield's approach to {{CLEF
  eHealth}} 2017 task 2. In: {{CEUR}} Workshop Proceedings: {{Working}} Notes
  of {{CLEF}} 2017: {{Conference}} and Labs of the Evaluation Forum (2017)

\bibitem{alshami2023harnessing}
Alshami, A., Elsayed, M., Ali, E., Eltoukhy, A.E., Zayed, T.: Harnessing the
  power of chatgpt for automating systematic review process: Methodology, case
  study, limitations, and future directions. Systems  \textbf{11}(7), ~351
  (2023)

\bibitem{anagnostou2017combining}
Anagnostou, A., Lagopoulos, A., Tsoumakas, G., Vlahavas, I.P.: Combining
  inter-review learning-to-rank and intra-review incremental training for title
  and abstract screening in systematic reviews. In: {{CEUR}} Workshop
  Proceedings: {{Working}} Notes of {{CLEF}} 2017: {{Conference}} and Labs of
  the Evaluation Forum (2017)

\bibitem{aum2021srbert}
Aum, S., Choe, S.: srbert: automatic article classification model for
  systematic review using bert. Systematic reviews  \textbf{10}(1), ~1--8
  (2021)

\bibitem{bramer2017optimal}
Bramer, W.M., Rethlefsen, M.L., Kleijnen, J., Franco, O.H.: Optimal database
  combinations for literature searches in systematic reviews: a prospective
  exploratory study. Systematic reviews  \textbf{6},  1--12 (2017)

\bibitem{callaghan2020statistical}
Callaghan, M.W., M{\"u}ller-Hansen, F.: Statistical stopping criteria for
  automated screening in systematic reviews. Systematic Reviews  \textbf{9}(1),
   1--14 (2020)

\bibitem{carvallo2020automatic}
Carvallo, A., Parra, D., Lobel, H., Soto, A.: Automatic document screening of
  medical literature using word and text embeddings in an active learning
  setting. Scientometrics  \textbf{125},  3047--3084 (2020)

\bibitem{carvallo2020neural}
Carvallo, A., Parra, D., Rada, G., Perez, D., Vasquez, J.I., Vergara, C.:
  Neural language models for text classification in evidence-based medicine.
  arXiv preprint arXiv:2012.00584  (2020)

\bibitem{chandler2019cochrane}
Chandler, J., Cumpston, M., Li, T., Page, M.J., Welch, V.A.: Cochrane Handbook
  for Systematic Reviews of Interventions. {John Wiley \& Sons} (2019)

\bibitem{chen2017ecnu}
Chen, J., Chen, S., Song, Y., Liu, H., Wang, Y., Hu, Q., He, L., Yang, Y.:
  {{ECNU}} at 2017 {{eHealth}} task 2: {{Technologically}} assisted reviews in
  empirical medicine. In: {{CEUR}} Workshop Proceedings: {{Working}} Notes of
  {{CLEF}} 2017: {{Conference}} and Labs of the Evaluation Forum (2017)

\bibitem{chiang2023vicuna}
Chiang, W.L., Li, Z., Lin, Z., Sheng, Y., Wu, Z., Zhang, H., Zheng, L., Zhuang,
  S., Zhuang, Y., Gonzalez, J.E., et~al.: Vicuna: An open-source chatbot
  impressing gpt-4 with 90\%* chatgpt quality. See https://vicuna. lmsys. org
  (accessed 14 April 2023)  (2023)

\bibitem{suhail2013methods}
Clark, J.: Systematic reviewing. In: Suhail A. R.~Doi, G.M.W. (ed.) Methods of
  Clinical Epidemiology. {Springer} (2013)

\bibitem{cohen2010prospective}
Cohen, A.M., Ambert, K., McDonagh, M.: A prospective evaluation of an automated
  classification system to support evidence-based medicine and systematic
  review. In: AMIA annual symposium proceedings. vol.~2010, p.~121. American
  Medical Informatics Association (2010)

\bibitem{cohen2006reducing}
Cohen, A., Hersh, W., Peterson, K., Yen, P.: Reducing workload in systematic
  review preparation using automated citation classification. Journal of the
  American Medical Informatics Association  \textbf{13}(2),  206--219 (2006)

\bibitem{collaboration2002cochrane}
Collaboration, C.: The cochrane library. Database available on disk and CDROM.
  Oxford, UK, Update Software  (2002)

\bibitem{crumley2005resources}
Crumley, E.T., Wiebe, N., Cramer, K., Klassen, T.P., Hartling, L.: Which
  resources should be used to identify rct/ccts for systematic reviews: a
  systematic review. BMC Medical Research Methodology  \textbf{5},  1--13
  (2005)

\bibitem{dettmers2023qlora}
Dettmers, T., Pagnoni, A., Holtzman, A., Zettlemoyer, L.: Qlora: Efficient
  finetuning of quantized llms. arXiv preprint arXiv:2305.14314  (2023)

\bibitem{di2017interactive}
Di~Nunzio, G.M., Beghini, F., Vezzani, F., Henrot, G.: An interactive
  two-dimensional approach to query aspects rewriting in systematic reviews.
  {{IMS}} unipd at {{CLEF eHealth}} task 2. In: {{CEUR}} Workshop Proceedings:
  {{Working}} Notes of {{CLEF}} 2017: {{Conference}} and Labs of the Evaluation
  Forum (2017)

\bibitem{di2018interactive}
Di~Nunzio, G.M., Ciuffreda, G., Vezzani, F.: Interactive sampling for
  systematic reviews. {{IMS}} unipd at {{CLEF}} 2018 {{eHealth}} task 2. In:
  {{CEUR}} Workshop Proceedings: {{Working}} Notes of {{CLEF}} 2018:
  {{Conference}} and Labs of the Evaluation Forum (2018)

\bibitem{kanoulas2017clef}
Kanoulas, E., Li, D., Azzopardi, L., Spijker, R.: {{CLEF}} 2017 technologically
  assisted reviews in empirical medicine overview. In: {{CEUR}} Workshop
  Proceedings: {{Working}} Notes of {{CLEF}} 2017: {{Conference}} and Labs of
  the Evaluation Forum (2017)

\bibitem{kanoulas2019clef}
Kanoulas, E., Li, D., Azzopardi, L., Spijker, R.: {{CLEF}} 2019 technology
  assisted reviews in empirical medicine overview. In: {{CEUR}} Workshop
  Proceedings: {{Working}} Notes of {{CLEF}} 2018: {{Conference}} and Labs of
  the Evaluation Forum. vol.~2380 (2019)

\bibitem{kanoulas2018clef}
Kanoulas, E., Spijker, R., Li, D., Azzopardi, L.: {{CLEF}} 2018 technology
  assisted reviews in empirical medicine overview. In: {{CEUR}} Workshop
  Proceedings: {{Working}} Notes of {{CLEF}} 2018: {{Conference}} and Labs of
  the Evaluation Forum (2018)

\bibitem{kopf2023openassistant}
K{\"o}pf, A., Kilcher, Y., von R{\"u}tte, D., Anagnostidis, S., Tam, Z.R.,
  Stevens, K., Barhoum, A., Duc, N.M., Stanley, O., Nagyfi, R., et~al.:
  Openassistant conversations--democratizing large language model alignment.
  arXiv preprint arXiv:2304.07327  (2023)

\bibitem{kozorovitsky2011identical}
Kozorovitsky, A.K., Kurland, O.: From``identical''to``similar'': Fusing
  retrieved lists based on inter-document similarities. Journal of Artificial
  Intelligence Research  \textbf{41},  267--296 (2011)

\bibitem{lagopoulos2018learning}
Lagopoulos, A., Anagnostou, A., Minas, A., Tsoumakas, G.: Learning-to-rank and
  relevance feedback for literature appraisal in empirical medicine. In:
  {{CEUR}} Workshop Proceedings: {{Working}} Notes of {{CLEF}} 2018:
  {{Conference}} and Labs of the Evaluation Forum. pp. 52--63. {Springer}
  (2018)

\bibitem{lee2018seed}
Lee, G.E., Sun, A.: Seed-driven document ranking for systematic reviews in
  evidence-based medicine. In: Proceedings of the 41st Annual International
  {{ACM SIGIR}} Conference on Research and Development in Information
  Retrieval. pp. 455--464 (2018)

\bibitem{lee2020biobert}
Lee, J., Yoon, W., Kim, S., Kim, D., Kim, S., So, C.H., Kang, J.: Biobert: a
  pre-trained biomedical language representation model for biomedical text
  mining. Bioinformatics  \textbf{36}(4),  1234--1240 (2020)

\bibitem{lu2021fantastically}
Lu, Y., Bartolo, M., Moore, A., Riedel, S., Stenetorp, P.: Fantastically
  ordered prompts and where to find them: Overcoming few-shot prompt order
  sensitivity. arXiv preprint arXiv:2104.08786  (2021)

\bibitem{minas2018aristotle}
Minas, A., Lagopoulos, A., Tsoumakas, G.: Aristotle university's approach to
  the technologically assisted reviews in empirical medicine task of the 2018
  {{CLEF eHealth}} lab. In: {{CEUR}} Workshop Proceedings: {{Working}} Notes of
  {{CLEF}} 2018: {{Conference}} and Labs of the Evaluation Forum (2018)

\bibitem{miwa2014reducing}
Miwa, M., Thomas, J., {O'Mara-Eves}, A., Ananiadou, S.: Reducing systematic
  review workload through certainty-based screening. Journal of Biomedical
  Informatics  \textbf{51},  242--253 (2014)

\bibitem{norman2019measuring}
Norman, C.R., Leeflang, M.M., Porcher, R., N{\'e}v{\'e}ol, A.: Measuring the
  impact of screening automation on meta-analyses of diagnostic test accuracy.
  Systematic reviews  \textbf{8}(1), ~243 (2019)

\bibitem{penedo2023refinedweb}
Penedo, G., Malartic, Q., Hesslow, D., Cojocaru, R., Cappelli, A., Alobeidli,
  H., Pannier, B., Almazrouei, E., Launay, J.: The refinedweb dataset for
  falcon llm: outperforming curated corpora with web data, and web data only.
  arXiv preprint arXiv:2306.01116  (2023)

\bibitem{robinson2023bio}
Robinson, A., Thorne, W., Wu, B.P., Pandor, A., Essat, M., Stevenson, M., Song,
  X.: Bio-sieve: Exploring instruction tuning large language models for
  systematic review automation. arXiv preprint arXiv:2308.06610  (2023)

\bibitem{scells2020clf}
Scells, H., Zuccon, G.: You can teach an old dog new tricks: {{Rank}} fusion
  applied to coordination level matching for ranking in systematic reviews. In:
  Proceedings of the 42nd European Conference on Information Retrieval. pp.
  399--414 (2020)

\bibitem{scells2017ltr}
Scells, H., Zuccon, G., Deacon, A., Koopman, B.: {{QUT}} ielab at {{CLEF
  eHealth}} 2017 technology assisted reviews track: {{Initial}} experiments
  with learning to rank. In: {{CEUR}} Workshop Proceedings: {{Working}} Notes
  of {{CLEF}} 2017: {{Conference}} and Labs of the Evaluation Forum (2017)

\bibitem{scells2019refinement}
Scells, H., Zuccon, G., Koopman, B.: Automatic boolean query refinement for
  systematic review literature search. In: Proceedings of the 28th World Wide
  Web Conference. pp. 1646--1656 (2019)

\bibitem{scells2020comparison}
Scells, H., Zuccon, G., Koopman, B.: A comparison of automatic boolean query
  formulation for systematic reviews. Information Retrieval Journal pp. 1--26
  (2020)

\bibitem{scells2020objective}
Scells, H., Zuccon, G., Koopman, B.: A computational approach for objectively
  derived systematic review search strategies. In: Proceedings of the 42nd
  European Conference on Information Retrieval. pp. 385--398 (2020)

\bibitem{scells2020conceptual}
Scells, H., Zuccon, G., Koopman, B., Clark, J.: Automatic boolean query
  formulation for systematic review literature search. In: Proceedings of the
  29th World Wide Web Conference. pp. 1071--1081 (2020)

\bibitem{singh2017iiit}
Singh, J., Thomas, L.: {{IIIT}}-{{H}} at {{CLEF eHealth}} 2017 task 2:
  {{Technologically}} assisted reviews in empirical medicine. In: {{CEUR}}
  Workshop Proceedings: {{Working}} Notes of {{CLEF}} 2017: {{Conference}} and
  Labs of the Evaluation Forum (2017)

\bibitem{EngeneAssessing2023}
Syriani, E., David, I., Kumar, G.: Assessing the ability of chatgpt to screen
  articles for systematic reviews. arXiv preprint arXiv:2307.06464  (07 2023)

\bibitem{alpaca}
Taori, R., Gulrajani, I., Zhang, T., Dubois, Y., Li, X., Guestrin, C., Liang,
  P., Hashimoto, T.B.: Stanford alpaca: An instruction-following llama model.
  \url{https://github.com/tatsu-lab/stanford_alpaca} (2023)

\bibitem{thomas2008methods}
Thomas, J., Harden, A.: Methods for the thematic synthesis of qualitative
  research in systematic reviews. BMC medical research methodology
  \textbf{8}(1), ~45 (2008)

\bibitem{touvron2023llama}
Touvron, H., Lavril, T., Izacard, G., Martinet, X., Lachaux, M.A., Lacroix, T.,
  Rozi{\`e}re, B., Goyal, N., Hambro, E., Azhar, F., et~al.: Llama: Open and
  efficient foundation language models. arXiv preprint arXiv:2302.13971  (2023)

\bibitem{touvron2023llama2}
Touvron, H., Martin, L., Stone, K., Albert, P., Almahairi, A., Babaei, Y.,
  Bashlykov, N., Batra, S., Bhargava, P., Bhosale, S., et~al.: Llama 2: Open
  foundation and fine-tuned chat models. arXiv preprint arXiv:2307.09288
  (2023)

\bibitem{wallace2012deploying}
Wallace, B.C., Small, K., Brodley, C.E., Lau, J., Trikalinos, T.A.: Deploying
  an interactive machine learning system in an evidence-based practice center:
  Abstrackr. In: Proceedings of the 2nd {{ACM}} International Health
  Informatics Symposium. pp. 819--824 (2012)

\bibitem{wallace2010semi}
Wallace, B.C., Trikalinos, T.A., Lau, J., Brodley, C., Schmid, C.H.:
  Semi-automated screening of biomedical citations for systematic reviews. BMC
  bioinformatics  \textbf{11}(1), ~55 (2010)

\bibitem{wang2021mesh}
Wang, S., Li, H., Scells, H., Locke, D., Zuccon, G.: Mesh term suggestion for
  systematic review literature search. In: Proceedings of the 25th Australasian
  Document Computing Symposium. pp.~1--8 (2021)

\bibitem{wang2023mesh}
Wang, S., Li, H., Zuccon, G.: Mesh suggester: A library and system for mesh
  term suggestion for systematic review boolean query construction. In:
  Proceedings of the Sixteenth ACM International Conference on Web Search and
  Data Mining. pp. 1176--1179 (2023)

\bibitem{wang2022little}
Wang, S., Scells, H., Clark, J., Koopman, B., Zuccon, G.: From little things
  big things grow: A collection with seed studies for medical systematic review
  literature search. In: Proceedings of the 45th International ACM SIGIR
  Conference on Research and Development in Information Retrieval. pp.
  3176--3186 (2022)

\bibitem{wang2022automated}
Wang, S., Scells, H., Koopman, B., Zuccon, G.: Automated mesh term suggestion
  for effective query formulation in systematic reviews literature search.
  Intelligent Systems with Applications p. 200141 (2022)

\bibitem{wang2022neural}
Wang, S., Scells, H., Koopman, B., Zuccon, G.: Neural rankers for effective
  screening prioritisation in medical systematic review literature search. In:
  Proceedings of the 26th Australasian Document Computing Symposium. pp. 1--10
  (2022)

\bibitem{ws2023chatgpt}
Wang, S., Scells, H., Koopman, B., Zuccon, G.: Can chatgpt write a good boolean
  query for systematic review literature search? In: Proceedings of the 46th
  International ACM SIGIR Conference on Research and Development in Information
  Retrieval. p. 1426–1436. SIGIR '23, Association for Computing Machinery,
  New York, NY, USA (2023). \doi{10.1145/3539618.3591703},
  \url{https://doi.org/10.1145/3539618.3591703}

\bibitem{wang2023generating}
Wang, S., Scells, H., Potthast, M., Koopman, B., Zuccon, G.: Generating natural
  language queries for more effective systematic review screening
  prioritisation. arXiv preprint arXiv:2309.05238  (2023)

\bibitem{wang2022self}
Wang, Y., Kordi, Y., Mishra, S., Liu, A., Smith, N.A., Khashabi, D.,
  Hajishirzi, H.: Self-instruct: Aligning language model with self generated
  instructions. arXiv preprint arXiv:2212.10560  (2022)

\bibitem{white2020pubmed}
White, J.: Pubmed 2.0. Medical reference services quarterly  \textbf{39}(4),
  382--387 (2020)

\bibitem{wu2018ecnu}
Wu, H., Wang, T., Chen, J., Chen, S., Hu, Q., He, L.: Ecnu at 2018 ehealth task
  2: {{Technologically}} assisted reviews in empirical medicine. Methods-a
  Companion to Methods in Enzymology  \textbf{4}(5), ~7 (2018)

\bibitem{xu2023qa}
Xu, Y., Xie, L., Gu, X., Chen, X., Chang, H., Zhang, H., Chen, Z., Zhang, X.,
  Tian, Q.: Qa-lora: Quantization-aware low-rank adaptation of large language
  models. arXiv preprint arXiv:2309.14717  (2023)

\bibitem{yang2023large}
Yang, C., Wang, X., Lu, Y., Liu, H., Le, Q.V., Zhou, D., Chen, X.: Large
  language models as optimizers. arXiv preprint arXiv:2309.03409  (2023)

\bibitem{yang2022goldilocks}
Yang, E., MacAvaney, S., Lewis, D.D., Frieder, O.: Goldilocks: Just-right
  tuning of bert for technology-assisted review. In: European Conference on
  Information Retrieval. pp. 502--517. Springer (2022)

\bibitem{zhang2023generation}
Zhang, R., Wang, Y.S., Yang, Y.: Generation-driven contrastive self-training
  for zero-shot text classification with instruction-tuned gpt. arXiv preprint
  arXiv:2304.11872  (2023)

\bibitem{zhao2021calibrate}
Zhao, Z., Wallace, E., Feng, S., Klein, D., Singh, S.: Calibrate before use:
  Improving few-shot performance of language models. In: International
  Conference on Machine Learning. pp. 12697--12706. PMLR (2021)

\bibitem{zou2018technology}
Zou, J., Li, D., Kanoulas, E.: Technology assisted reviews: {{Finding}} the
  last few relevant documents by asking {{Yes}}/{{No}} questions to reviewers.
  In: Proceedings of the 41st Annual International {{ACM SIGIR}} Conference on
  Research and Development in Information Retrieval. pp. 949--952 (2018)

\end{thebibliography}


\end{document}